\documentclass{ws-ijmpd}
\usepackage[super,compress]{cite}
\usepackage{aas_macros}
\usepackage{amsmath}
\usepackage{amsfonts}
\usepackage{amssymb}
\usepackage{graphicx}%
\usepackage{hyperref}

\begin{document}

\markboth{Gregory~V. Vereshchagin}
{Physics of non-dissipative ultrarelativistic photospheres}

\def\draftnote{\today\quad\currenttime\quad \qquad\jobname}%
%
\catchline{}{}{}{}{}
%

\title{\uppercase{Physics of non-dissipative ultrarelativistic photospheres}\footnote{Based on a talk presented at the Thirteenth Marcel Grossmann Meeting on General Relativity, Stockholm, July 2012.}}

\author{\uppercase{Gregory~V. Vereshchagin}}

\address{ICRANet, 65122, p.le della Repubblica, 10, Pescara, Italy and \\
ICRA and Dipartimento di Fisica, Universit\`a di Roma ``Sapienza'', \\
00185, p-le A. Moro 5, Rome, Italy\\
veresh@icra.it}

\maketitle

\begin{history} \received{18 November 2013} \revised{30 October 2013} \end{history}

\begin{abstract}
Recent observations, especially by the \emph{Fermi} satellite, point out the
importance of the thermal component in GRB spectra. This fact revives strong
interest in photospheric emission from relativistic outflows. Early studies
already suggested that the observed spectrum of photospheric emission from
relativistically moving objects  differs in shape from the Planck spectrum. However,
this component appears to be subdominant in many GRBs and the origin of the
dominant component is still unclear. One of the popular ideas is that energy
dissipation near the photosphere may produce a non-thermal spectrum and account
for such emission. Before considering such models, though, one has to
determine precise spectral and timing characteristics of the photospheric
emission in the simplest possible case. Hence this paper focuses on various
physical effects which make the photospheric emission spectrum different
from the black body spectrum and quantifies them.

\end{abstract}


\ccode{PACS numbers: 11.80.-m; 98.70.Rz}


\section{Introduction}

This paper intends to provide an overview of recent work focusing on the
photospheric emission in relativistic outflows. Such emission emerges when
relativistically expanding plasma becomes transparent to photons. It is
particularly relevant in the context of fireball (see e.g., Ref.~\refcite{1999PhR...314..575P}) and fireshell (see e.g., Ref.~\refcite{2007AIPC..910...55R}) models of
cosmic Gamma Ray Bursts (GRBs). This topic was discussed extensively
during the Thirteenth Marcel Grossmann Meeting in Stockholm in 2012, where a special
parallel session GRB1: Photospheric Emission in GRBs took place. In
addition some very recent progress on this topic will be discussed.

Gamma Ray Bursts are transient flashes of gamma radiation with duration
ranging from several milliseconds up to thousands of seconds. They are
characterized by strong variability down to a millisecond time scale of observed flux. Observed spectra are
nonthermal, and when integrated over time they are well fit by the
phenomenological Band model \cite{1993ApJ...413..281B} which can be thought of in terms of the
$\nu F_{\nu}(\nu)$ diagram, where $F_{\nu}$ is the net energy flux and $\nu$ is the frequency, as two power laws smoothly joined at the peak.

Recent observations, mostly by the \emph{Fermi} satellite, indicate that time resolved
spectra have a more complex shape than a simple Band one. In particular,
additional power law and black body components are required to obtain
satisfactory fits \cite{Vianello2013}. The presence of a black body component in
observed spectrum \cite{2013ApJ...770...32G} is of great importance, since it
points unambiguously towards the transition from an optically thick regime to an
optically thin one. Such a transition has long been expected in theoretical models of GRBs \cite{1986ApJ...308L..47G,1986ApJ...308L..43P}.

Gamma Ray Burst sources are assumed to be optically thick with a typical optical
depth on the order of $\tau\sim10^{15}$, see e.g., Ref.~\refcite{1999PhR...314..575P}.
An electron-positron plasma generated in the source and loaded with baryons is
expanding due to its radiative pressure
\cite{1986ApJ...308L..47G,1990ApJ...363..218P,2000A&A...359..855R}. At some
distance from the source the optical depth for Compton scattering decreases to
unity and most of the radiation trapped in the plasma gets released. The radial
distance from the source at which this transparency occurs is referred to as the
photospheric radius. The simplest spherically symmetric models of photospheric
emission predict the shape of the spectrum to be nearly a black body one
\cite{1986ApJ...308L..47G,1990ApJ...363..218P}. In contrast, the observed spectra
in GRBs look nonthermal, being significantly broader than the Planck one.
Since such simple photospheric models are at odds with the observations, attention was
turned to optically thin models of emission in GRBs. However, optically thin
models, in particular the ones based on synchrotron emission, were found to
contradict observations as well \cite{1998ApJ...506L..23P}.

Generally speaking, there are two effects which lead to the broadening of the
observed spectrum of photospheric emission with respect to the black body
shape, referred to as \textquotedblleft geometrical\textquotedblright\ and
\textquotedblleft physical\textquotedblright\ broadening, respectively \cite{Peer2013}.
\textquotedblleft Geometrical\textquotedblright\ broadening of the spectrum
occurs for several reasons. Firstly, the photosphere is not a sharp
surface, but is a region in space and time characterized by a probability for
photons to be scattered by electrons for the last time
\cite{2011ApJ...732...49P,2011ApJ...737...68B}: this is an analog of the last
scattering surface in cosmology, see e.g., Ref.~\refcite{2003moco.book.....D}. Secondly, the emission arriving at an
observer with a given arrival time originates from different parts of the
outflow, with different radial coordinaIn what follows we adopt thetes and angles. Hence the observed
spectrum represents a superposition of spectra produced at these different
parts of the outflow. \textquotedblleft Physical\textquotedblright\ broadening
results if dissipation of a part of the kinetic (or electromagnetic) energy into thermal energy occurs before
the plasma becomes transparent. Such dissipation may happen due to several
reasons: magnetic reconnection \cite{2006A&A...457..763G}, inelastic nuclear
collisions \cite{2010MNRAS.407.1033B} and shocks
\cite{2011MNRAS.415.1663T,2013ApJ...765..103L}. The generic dissipative photospheric model is based on
a simple idea: if the temperature of electrons is not equal to the temperature of
photons, Compton scattering produces distortions of the photon spectrum compared  to the Planck spectrum.
Photospheric models with dissipation are mostly concerned with the part of the
spectrum with energies higher than the peak energy. However, also the low energy part of the spectrum may be
modified due to Compton scattering, see Ref.~\refcite{2013MNRAS.tmpL.153A}.

One has to bear in mind that the detection of photospheric components in GRBs
provides unique information on the physical characteristics of the outflow,
and consequently on the properties of the source. In particular, constraints
on the Lorentz factor of the outflow \cite{2007ApJ...664L...1P} and on the radius at the
base of the outflow \cite{2013MNRAS.433.2739I} may be obtained.

In what follows we discuss various aspects of photospheric models in GRBs,
with particular emphasis on physical effects and implications for observations. In Section 2 the optical depth and the photospheric radius are introduced. In Section 3 radiative diffusion from relativistically expanding outflow is discussed. Section 4 considers the probability density of the last scattering of photons. Section 5 discusses the average number of scatterings in a finite relativistic outflow. In Section 6 the difference between the laboratory and arrival time pictures of the photosphere is illustrated. In Section 7 various methods for computation of observed light curves and spectra of the photospheric emission are presented. Section 8 gives a brief discussion of dissipative models of the photospheric emission. Discussion and conclusions follow.

\section{Optical depth}

We start from the definition of the key quantity:\ the \emph{photospheric
radius}. Recall that the optical depth along the light-like world line
${\mathcal{L}}$ is defined as \cite{1973rela.conf....1E}
\begin{equation}
\tau=\int_{\mathcal{L}}\sigma j_{\mu}dx^{\mu}, \label{tauWL}%
\end{equation}
where $\sigma$\ is the cross section, $j^{\mu}$ is the 4-current of particles, and
$dx^{\mu}$ is the coordinate length element of the world line.

Consider a spherically symmetric outflow with finite spatial extension expanding with ultrarelativistic velocity. Take a
light-like world line starting at time $t$ at the interior boundary $r=R$ of
the outflow and directed outwards. The optical depth given by equation
(\ref{tauWL}) is then (see e.g., Ref.~\refcite{1991ApJ...369..175A})
\begin{equation}
\tau=\int_{R}^{R+\Delta R}\!\!\!\sigma n\left(  1-\beta\cos\theta\right)
\frac{dr}{\cos\theta}, \label{tau}%
\end{equation}
where $r$ is radial coordinate and $\theta$ is the angle between the line of sight and photon trajectory, $\beta\simeq1-{1}/{2\Gamma^{2}}$, $r$ is used as a parameter along the
world line, $R+\Delta R$ is the radial coordinate at which the world line
crosses the outer boundary of the outflow, and $n$ is the number density of electrons measured in the laboratory reference frame (in which the source of the outflow is at rest).
If positrons are also present in the outflow,
their contribution must also be added. The quantity $\Delta R$ is found from
the equation of motion of the outflow.

For the computation of the optical depth in Eq.~(\ref{tau}) one has to know
density of electrons and Lorentz factor profiles along the light-like world
line. These quantities can be inferred, e.g., from relativistic hydrodynamic
simulations. Alternatively, one can assume certain profiles. In some cases the
integral (\ref{tau}) can be computed analytically:\ in particular this is the
case for a steady relativistic wind. In what follows we adopt the model of a
relativistic wind of finite duration with the radius $R_{0}$ at the radial position where the outflow originates (the base of the outflow), luminosity $L$, mass ejection rate $\dot{M}$ and laboratory width of the outflow $l$, see e.g., Ref.~\refcite{2002MNRAS.336.1271D}. Such a wind generated at the radius $R_0$ is initially energy-dominated and it expands with acceleration (accelerating phase). The acceleration terminates when the wind becomes matter-dominated and then expansion continues with constant velocity (coasting phase). Hence the laboratory electron density profile is%
\begin{equation}
n=\left\{
\begin{array}
[c]{cc}%
n_{0}\left(  \dfrac{R}{R_{0}}\right)  ^{-2}, & R(t)<R<R(t)+l,\\
\  & \\
0, & \mathrm{otherwise,}%
\end{array}
\right.
\end{equation}
where $n_{0}$\ is electron density at the base of the outflow. For the Lorentz
factor, following \cite{2011ApJ...733L..40M,2013ApJ...764...94V} we assume%
\begin{equation}
\Gamma(R)=\left\{
\begin{tabular}
[c]{lll}%
$\left(  \frac{R}{R_{0}}\right)  ^{a},$ & $R_{0}<R<\eta^{\frac{1}{a}}R_{0},$ &
$\text{accelerating phase}$\\
&  & \\
$\eta=\mathrm{const},$ & $R>\eta^{\frac{1}{a}}R_{0},$ & $\text{coasting
phase}$%
\end{tabular}
\right.
\end{equation}
where $\eta$\ accounts for the contribution of baryons%
\begin{equation}
\eta=L/\dot{M}c^{2},
\end{equation}
and $1/3\leq a\leq1$\ parametrizes the type of outflow. For a magnetically
dominated one we assume $a=1/3$, while for baryonic outflows $a=1$. Notice
that when $a=1$ and $l=R_{0}/c$ the thin shell approximation (see e.g., Ref.~\refcite{1990ApJ...365L..55S}) is recovered with baryonic loading parameter $B=\eta^{-1}$, total energy $E_{0}=Ll/c$, and total mass in
baryons $M=\dot{M}l/c$. This latter approximation is used within the fireshell
model of GRBs, see e.g., Ref.~\refcite{2007AIPC..910...55R} and references therein. In
contrast, the steady wind approximation used within the fireball model, see
e.g., Refs.~\refcite{2011ApJ...732...49P,2011ApJ...737...68B}, is obtained for $l\gg
R_{0}$.

\subsection{Photospheric radius}

The photospheric radius $R_{ph}$ is defined by equating expression (\ref{tau})
to unity and setting $\theta=0$. This radius may be obtained
analytically for the model introduced in the previous section. The result is
\begin{equation}
\frac{R_{ph}}{R_{0}}=\left\{
\begin{array}
[c]{cc}%
\left[  \frac{\tau_{0}}{2\left(  2a+1\right)  }\right]  ^{\frac{1}{2a+1}}, &
\tau_{0}\ll 2\eta^{2+\frac{1}{a}},\\
& \\
\dfrac{\tau_{0}}{2\eta^{2}}, & 2\eta^{2+\frac{1}{a}}\ll\tau_{0}\ll4\eta
^{4}\frac{l}{R_{0}},\\
& \\
\left(  \tau_{0}\frac{l}{R_{0}}\right)  ^{1/2}, & \tau_{0}\gg\eta^{\frac{2}%
{a}}\frac{R_{0}}{l}\quad\mathrm{and}\quad\tau_{0}\gg4\eta^{4}\frac{l}{R_{0}}.
\end{array}
\right.  \label{Rph}%
\end{equation}
where%
\begin{equation}
\tau_{0}=\sigma n_{0}R_{0}=\frac{\sigma E_{0}}{4\pi m_{p}c^{2}R_{0}l\eta
}=\frac{\sigma L}{4\pi m_{p}c^{3}R_{0}\eta}, \label{tau0}%
\end{equation}
and $m_{p}$ is the proton mass.

Recently a new classification of finite duration outflows with respect to photospheric emission has been proposed \cite{2013ApJ...772...11R}. Namely:

\begin{itemize}
\item \textbf{Photon thick outflows}, where along the light-like world line connecting
the origin and the observer, the number density in the outflow decreases
significantly. In this respect the outflow is \textquotedblleft a long
wind\textquotedblright, even if the laboratory thickness of the outflow may be small.

\item \textbf{Photon thin outflows}, where the number density of the outflow does
not change substantially along this light-like world line. In this respect the
outflow is \textquotedblleft a thin shell\textquotedblright\ even if the
duration of the explosion producing the plasma could be long.
\end{itemize}

For instance, a geometrically thin ultrarelativistically expanding shell may
be both thin or thick with respect to the photons propagating inside it:\ hence
the origin of the term.

For completeness consider also the case when the number density of
positrons exceeds the number density of baryons. Note that the electron-positron plasma in a GRB source (also in the presence of baryons) reaches thermal equilibrium before expansion
\cite{2008AIPC..966..191A} and it remains accelerating until it
becomes transparent to radiation. Due to the exponential dependence of the thermal
pair density on the temperature $T_{c}$, measured in the reference frame comoving with the outflow, and hence on the radial
coordinate, transparency is reached at $kT_{c}^{\pm}\simeq0.04m_{e}c^{2}$, where $k$ is the Boltzmann constant and $m_{e}$ is the electron mass, rather independent of initial conditions. Given the initial temperature
$T_{0}=\left(  {16\pi\sigma_{SB}R_{0}^{2}}/L\right)  ^{-1/4}$, where
$\sigma_{SB}$ is the Stefan-Boltzmann constant, and the adiabaticity of expansion
with $T_{c}=T_{0}R_{0}/r$ (see, e.g., Ref.~\refcite{2012arXiv1205.3512R}), we find
\begin{equation}
R_{ph}^{\pm}=\frac{1}{T_{c}^{\pm}}\left(  \dfrac{LR_{0}^{2}}{16\pi\sigma_{SB}%
}\right)  ^{1/4}. \label{rtrepwind}%
\end{equation}

There is a discussion in Ref.~\refcite{2013ApJ...772...11R}
which shows that even if these results were partially known, such a new
classification improves our physical understanding of finite outflows with
respect to the photospheric emission. In particular, it becomes clear that all
asymptotic expressions (\ref{Rph}) are relevant within both fireball and fireshell models.

All these results are derived assuming a  constant Lorentz factor across
the outflow. Hence the width of the shell remains constant during its
expansion. However, analytical (Refs.~\refcite{1990ApJ...365L..55S,1995PhRvD..52.4380B}) and numerical (Refs.~\refcite{1993MNRAS.263..861P,1993ApJ...415..181M}) hydrodynamic calculations show that a Lorentz factor gradient 
develops during the expansion of the outflow. This leads to its spreading at
sufficiently large radii in the coasting phase. Recently such hydrodynamic spreading was considered
in Ref.~\refcite{2014NewA...27...30R}.

In the coasting phase of expansion each differential shell within the outflow is expanding
with almost constant speed $v=\beta c\simeq c(1-1/2\Gamma^{2})$, so the
spreading is determined by the radial dependence of the Lorentz
factor $\Gamma(r)$. In a variable outflow there can be regions with $\Gamma(r)$
decreasing with radius and $\Gamma(r)$ increasing with radius. At sufficiently
large radii only the regions with increasing $\Gamma$ contribute to the
spreading.

From the equations of motion of the external and internal boundaries of this region we
obtain \cite{1993MNRAS.263..861P} the thickness of the region as a function of the
radial position of the region
\begin{equation}
l(R)=l+\frac{R}{2}\left(  \frac{1}{\Gamma_{i}^{2}}-\frac{1}{\Gamma_{e}^{2}%
}\right)  , \label{lR}%
\end{equation}
where $\Gamma_{e}$ and $\Gamma_{i}$ are the Lorentz factors at the external and
internal boundaries, $l$ is the width of the region at small $R$, and $R$
is the radial position of the inner boundary. Consider such a region in the
two limiting cases: when the relative Lorentz factor difference is strong, namely
$\Delta\Gamma=\Gamma_{e}-\Gamma_{i}\gtrsim\Gamma_{i}$, and when the relative Lorentz
factor difference is weak, namely $\Delta\Gamma=\Gamma_{e}-\Gamma_{i}\ll
\Gamma_{i}$.

In the former case the second term in parenthesis in equation~(\ref{lR}) can
be neglected, and we obtain that the spreading becomes efficient at
$R>R_{b}=2\Gamma_{i}^{2}l$, see Refs.~\refcite{1993ApJ...415..181M} and
\refcite{1993MNRAS.263..861P}. In the latter case we find the corresponding
critical radius of hydrodynamic spreading $R_{b}=(\Gamma_{i}/\Delta
\Gamma)\Gamma_{i}^{2}l\gg\Gamma_{i}^{2}l$. From Eq.~(\ref{lR}) one can
see that in both cases for $R\gg R_{b}$, the width of the outflow is increasing
linearly with radius $l(R)\simeq(\Delta\Gamma/\Gamma_{i})R/\Gamma_{i}^{2}$.

Hence the definition of the photospheric radius given above in Eq.~(\ref{Rph})
corresponds to the case of a weak Lorentz factor difference with $R_{b}\gg
R_{tr}$. Consider now the case of a strong relative Lorentz factor difference
across the outflow. Take an element of fluid with a constant number of
particles $dN$ in the part of the shell with a gradient of $\Gamma$. The internal
boundary of the element is moving with velocity $v$, and the external one is
moving with velocity $v+dv=v+\frac{dv}{dR}dr$, where $dR$ is the differential
thickness at some fixed laboratory time $t=0$ and the derivative ${dv}/{dR}$ is
taken at the same laboratory time. Then at time $t$ the width of the element
is $dl=dR+tdv$, its radial position is $R(t)=r_{0}+vt$, where $r_{0}$ is the
initial radial position of the element, and the corresponding laboratory density
is
\begin{equation}
n_{l}=\frac{dN}{dV}=\frac{dN}{4\pi R^{2}\left(  1+t\frac{dv}{dR}\right)
dr}=n_{0}\frac{r_{0}^{2}}{R^{2}\left(  1+t\frac{dv}{dR}\right)  },
\label{nspr}%
\end{equation}
where $\ n_{0}=\frac{dN}{dV_{0}}=\frac{dN}{4\pi r_{0}^{2}dR}$. In order to
compute the integral (\ref{tau}) one needs to find the expression for the baryonic
number density along the light-like world line. Taking into account
hydrodynamical spreading (\ref{lR}) one obtains
\begin{equation}
n=\frac{n_{0}}{\Gamma}\left(  \frac{R_{0}}{r}\right)  ^{2}\frac{1}{1+\frac
{2r}{\Gamma}\frac{d\Gamma}{dr}}, \label{nspralongtheray}%
\end{equation}
which is exact in the ultrarelativistic limit. An estimate for $d\Gamma/dr$ can be
given for a strong relative Lorentz factor difference $\Delta\Gamma\sim\Gamma$
in the outflow
\begin{equation}
\frac{d\Gamma}{dr}\sim\frac{\Delta\Gamma}{\Delta r}\sim\frac{1}{2\Gamma l},
\label{dGdr}%
\end{equation}
where $\Delta r\sim2\Gamma^{2}l$ is the distance inside the outflow along the
light-like world line. Integrating expression~(\ref{dGdr}) the Lorentz
factor dependence on the radial coordinate along this light-like world line is obtained
\begin{equation}
\Gamma(r)\sim\sqrt{\frac{r-R}{l}}. \label{Gammar}%
\end{equation}
Since we are interested in the asymptotics when hydrodynamic spreading is
essential, one can assume that $r\gg R$ in the integral (\ref{tau}), which is
equivalent to $R\ll2\Gamma^{2}l$.\ Under this condition the integral can be
performed analytically and the photospheric radius is obtained as
\begin{equation}
\frac{R_{ph}}{R_{0}}=\left(  \frac{\tau_{0}}{8}\frac{l}{R_{0}}\right)  ^{1/2}.
\label{tauspreading}%
\end{equation}
This result coincides with the last line in Eq.~(\ref{Rph}) up to a numerical
factor. However, its physical meaning is different: it represents the photon thick
asymptotics of Eq.~(\ref{tau}), since $R_{ph}\ll2\Gamma^{2}l$.

Formulas (\ref{Rph}), (\ref{rtrepwind}) and (\ref{tauspreading}) represent
asymptotic expressions for the photospheric radius in different physical
situations relevant for GRBs. These results are obtained within the simple
model of relativistic wind with finite duration, while the last expression
accounts also for radial spreading of the outflow. It is essential that these
equations are supplied with the corresponding ranges of validity. In realistic
cases with hydrodynamic profiles obtained by e.g., numerical simulations, one
has to perform numerical integration of Eq.~(\ref{tau}) with $\theta=0$ and
find the photospheric radius by equating the value obtained by such an
integration to unity.

\subsection{Shape of the photosphere}

Similarly, the shape of the photosphere can be recovered by equating the
integral (\ref{tau}) to unity without requiring $\theta=0$. In such a case one
obtains a surface $R_{ph}(\theta)$. The
shape of this surface at a given laboratory time is far from being a sphere:
it is concave for photon thick outflows and convex for photon thin ones. It is
instructive to consider the limiting cases of a steady wind on the one hand, and an
infinitely thin shell on the other hand.

\begin{figure}[ptb]
\centering\includegraphics[width=3.3in]{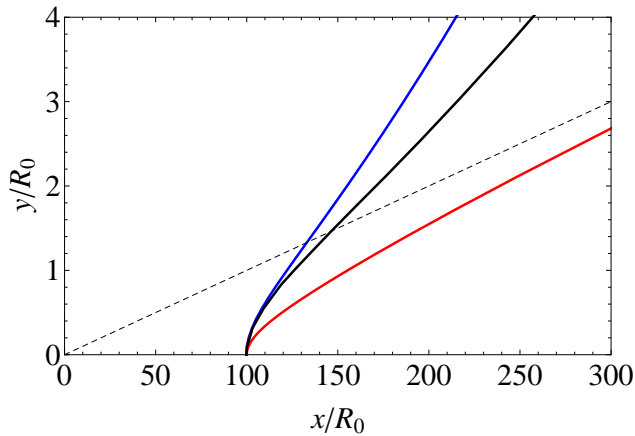}
\caption{The shape of photospheres
of a steady coasting relativistic wind (blue curve) and a steady accelerating
relativistic wind (black curve) as well as the time-integrated photosphere of an
infinitely thin shell (red curve) for $\Gamma=100$. The dashed line shows the
relativistic beaming angle.}%
\label{WindShell}%
\end{figure}

Firstly, the photosphere of a steady relativistic wind with $\Gamma={}$const
analyzed in Ref.~\refcite{1991ApJ...369..175A} is given by%
\begin{equation}
\frac{r}{R_{0}}=\tau_{0}\left(  \frac{\theta}{\sin\theta}-1+\frac{1}{2\eta
^{2}}\right)  , \label{PhotosphereInfWindCoasting}%
\end{equation}
which is a static surface having a concave shape, see Fig.~\ref{WindShell}.
Secondly, the photosphere of a steady accelerating relativistic wind
is described by a cubic equation. It corresponds to a static surface with
curvature larger than that of the coasting wind, see Fig.~\ref{WindShell}.
Thirdly, the photosphere of an infinitely thin shell\footnote{An infinitely thin
shell is understood as the limit $l\rightarrow0$ with $n_{0}l=$const.} at a
fixed laboratory time within the relativistic beaming cone $\theta
=\arccos\left(  1-\frac{1}{2\eta^{2}}\right)  $ is an infinitely thin ring.
The collection of such rings for all laboratory times represents a surface
given by
\begin{equation}
\frac{r}{R_{0}}=\left[  {\tau_{0}\frac{1-\left(  1-\frac{1}{2\eta^{2}}\right)
\cos\theta}{|\cos\theta-\left(  1-\frac{1}{2\eta^{2}}\right)  |}}\right]
^{1/2}. \label{ShellPhotosphere}%
\end{equation}
The curvature of this surface is even larger than that of an accelerating wind,
as can be seen from Fig.~\ref{WindShell}. This surface represents the
asymptotic limit of photospheres of photon thin outflows. The surfaces (\ref{PhotosphereInfWindCoasting}) and
(\ref{ShellPhotosphere}) give the position of the corresponding photon thick
and photon thin outflow photospheres with very good accuracy.

\section{Radiative diffusion}

The definition of the photosphere implies that at this position in space the
outflow as a whole becomes transparent to radiation. However, emission emerges from the
outflow when it is optically thick as well. Such emission is due to radiative
diffusion which transfers the energy from deeper parts of the outflow towards
its surface. Naively one can think that such an effect is negligible in
ultrarelativistic outflows. However, this is not the case. This effect,
usually neglected in the literature based on considerations of steady winds,
plays a crucial role in photon thin outflows \cite{2013ApJ...772...11R}.
Actually, near the photosphere, photons are not produced in the outflow since
three-particle interactions such as bremsstrahlung and double Compton
scattering become inefficient (freeze out) at smaller radii. Hence photons may diffuse from the interior of the outflow to the boundary and eventually escape before the outflow reaches the photospheric radius. In such a
case there will be no emission at the photospheric radius, since most photons
have already escaped. This is exactly the case for photon thin outflows, as we will
see below.

Consider the diffusion crossing time on which a photon is expected to cross the
outflow with comoving radial thickness $l_{c}=\Gamma l$. This time is given by $t_{D,c}=l_{c}%
^{2}/D_{c}$, where the diffusion coefficient is $D_{c}=(c\lambda_{c})/3=c/(3\sigma
n_{c})$ and $\lambda_{c}$ and $n_{c}$\ are the comoving mean free path of
photons and the comoving electron number density, respectively. One may find the
radial position $R_{D}$ at which the outflow arrives by the time $t_{D,c}$. At
this time, measured from the beginning of the expansion, photons actually cross the
entire width of the outflow by diffusion. Neglecting the initial brief
acceleration phase when diffusion is irrelevant, from the equation of motion
of the outflow $R=\beta ct\simeq\Gamma ct_{c}$, where $t_{c}$ is time measured
in comoving frame, and taking into account Eq.~(\ref{tau0}) we obtain
\begin{equation}
R_{D}=\left(  \tau_{0}\eta^{2}R_{0}l^{2}\right)  ^{1/3}. \label{diffradius}%
\end{equation}
This diffusion radius should be compared to the photospheric radius (\ref{Rph}).
From the last line in Eq.~(\ref{Rph}) one finds%
\begin{equation}
\frac{R_{D}}{R_{ph}}=\left(  \frac{\eta^{4}l}{\tau_{0}R_{0}}\right)
^{\frac{1}{6}}, \label{rdrph}%
\end{equation}
which is valid for any acceleration model of the outflow.

For photon thick outflows one has $4\eta^{4}l\gg\tau_{0}R_{0}$ and the
diffusion radius is always larger than the photospheric radius. In other
words, the width of the outflow is so large that photons have no time to
diffuse out by the moment the outflow becomes transparent. Hence diffusion is
irrelevant for photon thick outflows.

The situation is the opposite for photon thin outflows, when $\tau_{0}R_{0}%
\gg4\eta^{4}l$ and the diffusion radius is always smaller than the
photospheric radius. This implies that in photon thin outflows radiation
decouples not at the photospheric radius, defined by Eq.~(\ref{Rph}), but
at the diffusion radius defined by Eq.~(\ref{diffradius}), when the expanding
plasma is still opaque. In this sense the characteristic radius of the
photospheric emission from photon thin outflows is the diffusion radius.

\section{Probability density}

One has to keep in mind that photons decouple from the plasma not only at a
surface given for photon thick outflows by the condition $\tau=1$. The last
scattering of photons occurs in a finite region of spacetime near this surface.
Hence the photosphere is not a sharp surface, but it is ``fuzzy"
\cite{2008ApJ...682..463P,2011ApJ...737...68B}. Since the outflow is dynamical
and finite in spatial extension, the finiteness effects are expected to play an
important role for the formation of the photospheric emission. Such effects have been
recently investigated \cite{2013ApJ...767..139B} by means of Monte Carlo
simulations using the model of a relativistic wind with finite duration and
considering both coherent and Compton scattering of photons.

Most of energy reaching an observer is emitted from the region near the
photosphere, where the probability density function along the ray reaches its
maximum. This function \cite{2013ApJ...772...11R} is given by
\begin{equation}
P(r,\theta,t)\propto\frac{d}{ds}\exp[-\tau(r,\theta,t)], \label{PDF}%
\end{equation}
where $s$ is the distance measured along the light-like world line. When the
time dependence in this equation is discarded, this $P(r,\theta)$ coincides
with the probability density function of the last scattering defined in
Ref.~\refcite{2008ApJ...682..463P}.

\begin{figure}[ptb]
\centering%
\begin{tabular}
[c]{c}%
\includegraphics[width=3.3in]{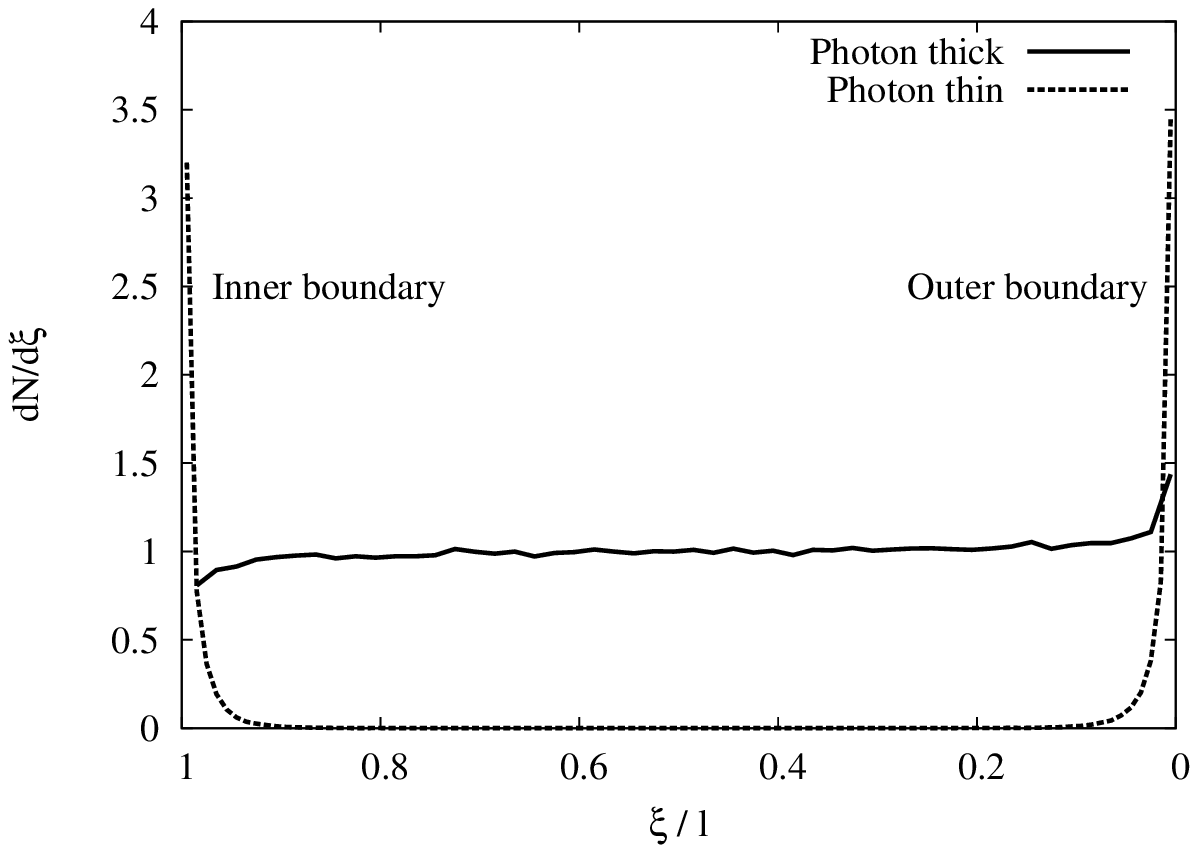}\\
\includegraphics[width=3.3in]{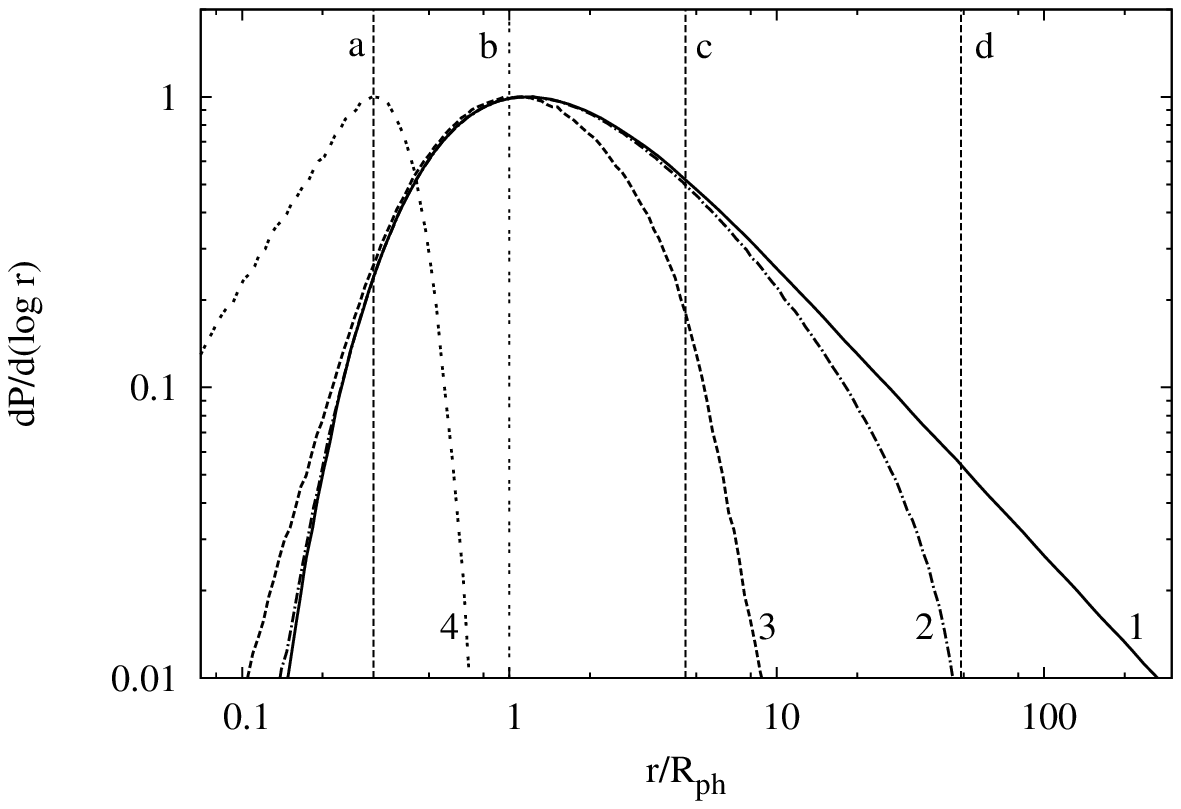}
\end{tabular}
\caption{Upper panel: Probability density of last scattering as a function of the
normalized depth for photon thick and photon thin outflows (in the latter case
decreased by a factor of 10 for a better presentation). \newline Lower panel:
probability density function for the position of last scattering in the
following cases: infinite and steady wind (1), photon thick case with
$\Gamma=500$ (2), photon thick case with a smaller Lorentz factor $\Gamma=300$
(3), and photon thin outflow (4). The vertical line (a) represents the
diffusion radius in the photon thin case, line (b) represents the photospheric
radius while lines (c) and (d) show the transition radius $R_{t}=\eta R_{0}$
for curves (2) and (3), respectively. \newline Reproduced from Ref.~\protect\refcite{2013ApJ...767..139B}.}%
\label{Fig:proba}%
\end{figure}

The probability density of the photon last scattering position is shown in 
Fig.~\ref{Fig:proba} as a function of the depth $\xi$\ measured from the outer boundary
of the outflow (top panel) and as a function of the radial coordinate (bottom
panel). In photon thick outflows the photon decoupling is local and almost
independent of $\xi$ (solid curve in the top panel of Fig.~\ref{Fig:proba}).
Instead, this probability density depends strongly on the radial coordinate
(curves 1--3 in the bottom panel of \ref{Fig:proba}). The probability density
function of last scattering is found to be close to the one of an infinite
steady wind studied in Ref.~\refcite{2011ApJ...737...68B}. The difference emerges due
to the presence of boundaries, and it is manifested in an exponential cut off in
the probability density at large radii.

In photon thin outflows there is enough time for photons to be transported to 
the boundaries by diffusion as discussed above, see also
Ref.~\refcite{2013ApJ...772...11R}. As a result the probability density as a function of
the depth peaks at the boundaries (dashed curve in the top panel of Fig.~\ref{Fig:proba}).

It is also clear (see curve 4 in the bottom panel of
\ref{Fig:proba}) that most photons escape from the photon thin outflow well
before the photospheric radius, namely near diffusion radius defined in Eq.~(\ref{diffradius}). 
This result confirms that radiation diffusion plays an
important role for this type of outflow. It actually determines the shape
of both the instantaneous and time integrated spectra as seen by a distant
observer, as well as the light curve \cite{2013ApJ...772...11R}.

\section{Average number of scatterings}

Another difference between the photon thick and photon thin cases manifests itself in the
average number of scatterings as a function of the initial optical depth
\cite{2013ApJ...767..139B}. The average number of scatterings is defined as
\begin{equation}
\langle N\rangle=\int_{t_{i}}^{t_{f}}\frac{cdt_{c}}{\lambda_{c}}=\int_{t_{i}%
}^{t_{f}}\sigma\langle n_{c}\rangle cdt_{c}, \label{int}%
\end{equation}
where $t_{c}$ is the comoving time, $\lambda=1/(\sigma_{T}n_{c})$ is the comoving
mean free path of photons, $n_{c}$ is the comoving density, $t_{i}$ is the initial
comoving time, and $t_{f}$ is the final comoving time when the photon leaves the
outflow. The integral (\ref{int}) is taken along the average photon path. In an
optically thick medium when diffusion is neglected, this path is given by the
equation of motion of the outflow
\begin{equation}
r=r_{i}+\beta ct, \label{avphpath}%
\end{equation}
where $r_{i}$ is the laboratory radial position of the photon at the initial
laboratory time. In the photon thick case photons stay in the outflow long
after decoupling, so one can set $t_{f}\rightarrow\infty$. Then, taking into account
relations $n_{c}=n_{l}/\Gamma$, $dr=\beta cdt$, and $t_{c}=t/\Gamma$ along the
world line (\ref{avphpath}), we obtain
\begin{equation}
\langle N\rangle=\int_{t_{i}}^{\infty}\sigma\frac{n}{\Gamma}\frac{cdt}{\Gamma
}=2\tau_{i},
\end{equation}
where $\tau_{i}$ is optical depth of the outflow at $r_{i}$.

In the photon thin case the photons decouple from the outflow near its boundaries, and the
time interval needed for the photon to reach them by a random walk can be
estimated as $t^{c}_{D}=l_{c}^{2}/D_{c}$, where $l_{c}=\Gamma l$ is the
comoving radial thickness of the outflow, and the diffusion coefficient is
$D_{c}=(c\lambda_{c})/3=c/(3\sigma n_{c})$. When this time is much less than
the characteristic time of expansion, equal to $t_{i}$, which is the case when the
initial radius $r_{i}$ is much larger than radius of diffusion $R_{D}$, we
have
\begin{equation}
\langle N \rangle\simeq3\tau_{i}^{2},
\end{equation}
In the opposite case when initial radius $r_{i}$ is much smaller than the
radius of diffusion $R_{D}$, we have
\begin{equation}
\langle N\rangle\simeq\frac{1}{\Gamma^{2}}\sqrt{\tau_{i}\tau_{0}\frac{R_{0}%
}{l}}.
\end{equation}

\begin{figure}[ptb]
\centering
\includegraphics[width=3.3in]{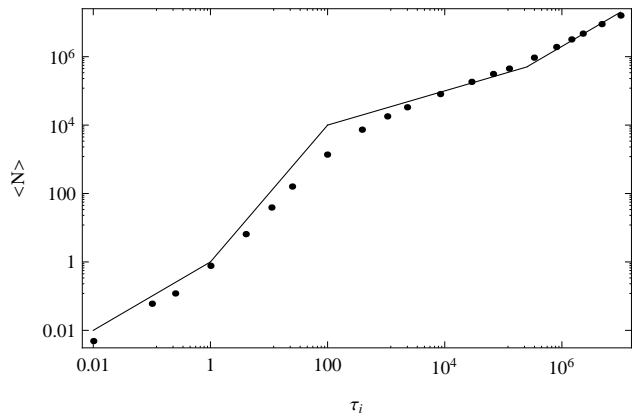} \caption{Average number of scatterings
as function of initial optical depth $\tau_{i}$. Results of Monte Carlo simulations are
shown with points, while analytic results are shown by the corresponding lines. In the optically thin regime
the number of scatterings is proportional to $\tau_{i}$. For $1<\tau_{i}%
<\tau(R_{D})$ it is proportional to $\tau_{i}^{2}$. For $\tau(R_{D})<\tau
_{i}<\tau(R_{t})$ it is proportional to $\tau_{i}^{1/2}$. For even larger
$\tau_{i}$ corresponding to the photon thick asymptotic the number of
scatterings is again proportional to $\tau_{i}$. In these simulations $\Gamma=100$, $l=10^{8}$ cm and $\tau_{0}=10^{14}$. Reproduced from Ref.~\protect\refcite{2013ApJ...767..139B}.}%
\label{Fig:num_scatt}%
\end{figure}
The average number of scatterings in all these cases is shown in Fig.~\ref{Fig:num_scatt}, 
together with the results of Monte Carlo simulations. Indeed,
when the outflow is photon thick we have $\langle N\rangle\propto\tau_{i}$.
This result is in contrast with a static optically thick finite medium \cite{1979rpa..book.....R} where $\langle
N\rangle\propto\tau_{i}^{2}$. However, in photon thin outflows the number of
scatterings is increasing with $\tau_{i}$ even more slowly, namely as $\langle
N\rangle\propto\tau_{i}^{1/2}$\ for most of the photons. Instead, for those
photons which scatter at sufficiently large radii, larger than the
diffusion radius, we have $\langle N\rangle\propto\tau_{i}^{2}$ as in the static medium. These photons
arrive in the tail of the light curve.

These results have important implications in dissipative models of GRBs where
either kinetic or electromagnetic energy gets converted into thermal energy of the outflow when it is still optically
thick. They imply in particular that in contrast with a static medium, in a
relativistically moving medium the spectral distortion is harder to achieve,
since the number of scatterings is less in the latter case.

\section{Laboratory time versus arrival time pictures}

Once the photospheric radius (or diffusion radius) is known such averaged properties of the photospheric emission as the peak energy in the observed spectrum and the duration of emission can be determined. In order to know the detailed light curve and time resolved spectra one has to define the notion of a dynamic photosphere. Such a dynamic photosphere is determined in general by the condition $\tau(r,t,\theta)=1$. For an outflow with finite spatial extension this dynamic photosphere is located inside it and typically crosses the outflow from the outer to the inner boundary while the total optical depth of the outflow defined by Eq.~(\ref{tau}) decreases to unity.

It is well known that the optical depth is a Lorentz invariant quantity, for
relevant discussion see Ref.~\refcite{Siutsou2013}. However, due to relativistic
motion the picture of the dynamic photosphere in relativistic outflow looks
different in the laboratory and observer reference frames. A distant observer uses the
arrival time $t_{a}$\ which is related to the laboratory time $t$, the radial coordinate $r$ and the
angle $\theta$ measured in the\ laboratory frame as%
\begin{equation}
t_{a}=t-r\cos\theta/c. \label{arrtime}%
\end{equation}
Hence photons emitted at one and the same laboratory time but from points with
different $\theta$ are detected at different arrival times; vice versa,
photons detected at the same values of the arrival time have been emitted at
different laboratory times and from points with different $\theta$. At a given
arrival time one has an \textquotedblleft equitemporal
surface\textquotedblright, see e.g., Ref.~\refcite{2001A&A...368..377B}, i.e., the
surface locus of points emitting photons with the same value of the arrival
time $t_{a}$. The corresponding equitemporal surface of the photospheric emission
is referred to as the PhE surface\cite{2013ApJ...772...11R}.

Given the hydrodynamic profiles of the number density of electrons $n(r,t)$
and the Lorentz factor $\Gamma(r,t)$ of the spherically symmetric outflow, one can
compute the optical depth and find the surface corresponding to the condition
$\tau(r,t,\theta)=1$. For the simplest case of a relativistic wind with finite duration
there is only one such surface at a given laboratory time $t$. In general,
however, for a single peaked density profile there are two surfaces:\ the
outer photosphere and the inner one, see e.g., Ref.~\refcite{Begue2013}. The inner
photosphere appears because of two opposite effects:\ the increase of density
with decreasing depth, and the decrease of density due to expansion. For the
outer photosphere both effects work in the same direction.

The distant observer, however, can see only one photosphere at a given arrival
time, namely the PhE surface. For a single peaked density profile the light curve seen
by this observer will be also single peaked. Emission originating from the
outer photosphere arrives first. The peak in the light curve corresponds
roughly to emission from the peak in the laboratory density profile. Emission from the
inner photosphere arrives in the tail of the light curve.

For more complex hydrodynamic profiles with multiple peaks in the electron number
density, multiple photospheres may appear, which may result in a spiky light
curve of the photospheric emission with a variable time scale down to $R_0/c$, i.e., milliseconds \cite{2005ApJ...628..847R}. In this respect observations may be used
for reconstruction of the underlying hydrodynamic profiles, see
Ref.~\refcite{2013MNRAS.433.2739I}. In addition, departure from spherical symmetry
manifested in the appearance of an angular dependence of hydrodynamic quantities, in
particular of the bulk Lorentz factor, is also crucial for the formation of
observed spectra of photospheric emission \cite{2013MNRAS.428.2430L}.

\section{Spectra and light curves as seen by a distant observer}

\begin{figure}[ptb]
\centering
\includegraphics[width=3.3in]{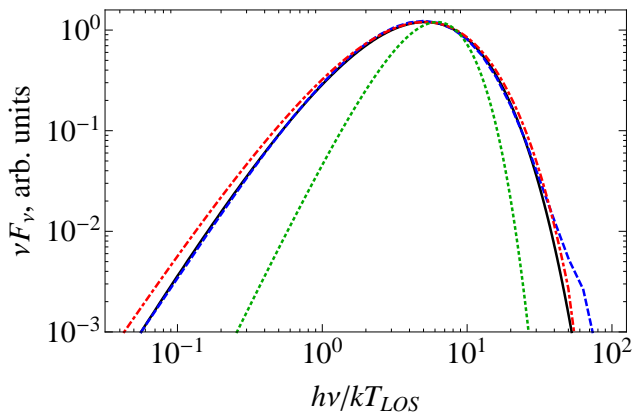}
\caption{The spectrum of photospheric
emission from photon thick outflow obtained with different approximations, but with the same parameters of the outflow.
The dotted curve shows the Planck spectrum. 
The dashed-dotted curve shows the result
from Ref.~\protect\refcite{2013ApJ...772...11R} obtained using the fuzzy photosphere
approximation. 
The dotted curve shows the result from Ref.~\protect\refcite{2013ApJ...767..139B}
obtained from Monte Carlo simulations. 
The solid curve shows the result from
Ref.~\protect\refcite{2013MNRAS.tmpL.153A} obtained by the solution of the radiative transfer
equation with the Fokker-Planck approximation to the collision integral.}
\label{photonthick}%
\end{figure}

Several methods of computing the spectra and light curves of photospheric
emission as seen by a distant observer have been proposed, in particular:
integration over the PhE surface \cite{2011ApJ...732...26M,2013ApJ...772...11R};
integration over volume with attenuation factors
\cite{2011ApJ...732...49P,2013MNRAS.428.2430L}; approximations to the
radiative transfer \cite{2011ApJ...737...68B,2013ApJ...772...11R}; Monte Carlo
simulations of photon scattering
\cite{2010MNRAS.407.1033B,2012MNRAS.422.3092G,2013ApJ...767..139B,2013arXiv1306.4822I}%
; Fokker-Planck approximation to the collision integral with an anisotropic photon field (generalized Kompaneets equation) \cite{2013MNRAS.tmpL.153A}
and relativistic Boltzmann equations \cite{2005ApJ...628..857P,Benedetti2013}.
It is remarkable that several very different methods produce practically
the same results, see Fig.~\ref{photonthick}. 

All these results indicate that the photospheric spectrum is wider than the
Planck one. There is an increase in the low energy slope of the
spectrum, resulting in the photon index $\alpha\simeq0$, in contrast with the
Planck photon index $\alpha=1$.

The computation method based on the \emph{radiative transfer equation} turns
out to be useful since it represents a second step beyond the simplest approach
adopted by Goodman \cite{1986ApJ...308L..47G}. In fact, for estimating the
observed spectrum Goodman used superpositions of Planck spectra corresponding
to different emitting regions with temperatures obtained from hydrodynamic
equations. In approximate solutions of the radiative transfer equations using
``fuzzy photosphere" and ``sharp photosphere" approximations \cite{2013ApJ...772...11R} the spectrum
coming from a given region no longer has the Planck shape, although the source function
in the comoving frame is still assumed to be isotropic and thermal. Even if
the approximations adopted in this approach are plausible and consistent,
they must be verified by different methods.

\emph{Monte Carlo simulation} is a completely independent method. Here each
photon is followed in the laboratory reference frame where the plasma is initially at rest while it
experiences numerous collisions with the cross sections of Compton and
isotropic scattering models, until it ceases scattering. Photons are injected
in the expanding plasma well before it becomes transparent.The resulting photons
constitute the final spectrum. The drawbacks of this approach are:\ a
prescribed distribution of an electron component and the impossibility to account for the
stimulated emission of photons. The possibility to take into account Pauli blocking in Monte Carlo simulations of degenerate relativistic plasmas has  recently been discussed in Ref.~\refcite{2013JCoPh.249...13T}. The first limitation originates from the fact
that Monte Carlo simulations need a prescribed background of electrons, and
any back-reaction of photons on the electron distribution can be only accounted for
by an iterative scheme. The second limitation constrains the spectrum of photons
in the optically thick region to have a Wien shape instead of the Planck one, if
the model of Compton scattering is used. In addition, the good statistics required
to resolve both the low and high energy parts of the photon spectrum imply the need
for a large number of photons which in turn demands long computational times.

The \emph{Fokker-Planck approximation} to the Boltzmann equation allows one to take into account the stimulated emission of photons. However, it does  not allow accounting for variations in the distribution function of the electron component. In this approach
which solves partial differential equations, a rather good resolution in the
spectrum can be achieved. In the recent study \cite{2013MNRAS.tmpL.153A} the formation of the spectrum from the photospheric emission in relativistic winds is analyzed with this method. The comoving temperature is assumed to depend on the laboratory radius as $T\propto r^{-k}$, where $k=2/3$ corresponds to the law of adiabatic expansion of the relativistic coasting wind. The low energy power law index $\alpha$ of the observed spectrum is found to be a function of the power law index $k$ in the temperature distribution as $\alpha\simeq 1-k$.

Finally, numerical solution of the \emph{system of Boltzmann equations} for
electrons and photons is the most promising method. Here Compton scattering of
photons is followed from high optical depth regions to low optical depth ones,
and the complete evolution of both the photon and electron distributions can be
obtained self-consistently. Since finite difference methods are involved, the
only limitation in this approach is the size of the grid in the phase space.

Next we give a short overview of methods suggested in
Ref.~\refcite{2013ApJ...772...11R}. 
The basis of the spectrum and flux calculation is the radiative transfer equation for
the specific intensity $I_{\nu}$ along the ray (see e.g.,
Ref.~\refcite{1979rpa..book.....R})
\begin{equation}
\frac{dI_{\nu}}{ds}=j_{\nu}-\kappa_{\nu}I_{\nu}, \label{radtransf}%
\end{equation}
where $j_{\nu}$ is the monochromatic emission coefficient for frequency $\nu$,
$\kappa_{\nu}$ is the absorption coefficient and $s$ is the distance measured along
the ray.

The spectral intensity of radiation measured at infinity from a ray coming to an observer at
some arrival time $t_{a}$ is given by the formal solution of this equation
\cite{2011ApJ...737...68B}
\begin{align}
I_{\nu}(\nu,\rho,t_{a})  & =\int\mathcal{I}_{\nu}(\nu,r,\theta,t)\frac{d}%
{ds}\left\{  \exp[-\tau(\nu,r,\theta,t)]\right\}  \,ds\\
& =\int\mathcal{I}_{\nu}(\nu,r,\theta,t)\exp[-\tau(\nu,r,\theta,t)]\,d\tau, \nonumber
\end{align}
where $\mathcal{I}_{\nu}(r,\theta,t)$ is the source function, equal to the
ratio of emission and absorption coefficients $\mathcal{I}_{\nu}=j_{\nu
}/\kappa_{\nu}$, the optical depth $\tau$ is given by
\begin{equation}
\tau=\int_{s}^{\infty}\kappa_{\nu}ds,\label{taukappa}%
\end{equation}
coinciding with Eq.~(\ref{tauWL}), and the variables $(r,\theta,t)$ are
connected by Eq.~(\ref{arrtime}) and $\rho=r\sin\theta$.

The total observed flux is an integral over all the rays
\begin{equation}
\label{Fnu}F_{\nu}(\nu,t_{a})=2\pi\Delta\Omega\int I_{\nu}(\nu,\rho,t_{a})\,
\rho\, d\rho\, ,
\end{equation}
where $\Delta\Omega$ is the solid angle of the observer's detector as seen
from the outflow in the laboratory frame and $2\pi\rho d \rho$ is an element
of area in the plane of the sky.

The emissivity $j_{\nu}$ is assumed to be thermal and isotropic in the comoving
frame of the outflow and the comoving opacity is $\kappa_{\nu,c}=\mathrm{const}%
$. The laboratory source function is then%
\begin{equation}
\mathcal{I}_{\nu}(\nu,r,\theta,t)=\frac{2h}{c^{2}}\frac{\nu^{3}}{\exp\left(
\frac{h\nu\Gamma(1-\beta\cos\theta)}{kT_{c}(r,t)}\right)  -1},
\end{equation}
where $h$ is Planck's constant. This approximation is justified when the
radiation field is tightly coupled to the matter. The photospheric emission
comes from the entire volume of the outflow, \cite{2001A&A...368..377B} and the
computational method sketched above is closely related to that used in
Ref.~\refcite{2011ApJ...737...68B}, where the concept of the \textquotedblleft fuzzy
photosphere\textquotedblright\ was introduced. This method is referred to as the
\emph{fuzzy photosphere} approximation.

Most of the energy reaching the observer is emitted from the region near the PhE surface,
where the probability density function along the ray given by Eq.~(\ref{PDF})
reaches its maximum. For this reason the dynamics of the PhE surface discussed in the
previous section determines both the light curves and spectra of the observed
photospheric emission. Assuming that all the energy
comes from the PhE surface only, i.e., a surface instead of the volume discussed above,
the computation may be reduced to a one-dimensional integration by substitution
of the function $P$ by a Dirac delta function. Such a cruder approximation, in
contrast with the fuzzy photosphere one, is referred to as the \emph{sharp
photosphere} approximation.

\section{Dissipative photospheres}

In spite of the title this review, I conclude with a brief discussion of dissipative photospheric
models of GRBs. All these models assume that some energy is eventually
converted into photons in the optically thick regime, which makes the
photospheric component brighter.\ In the literature such dissipation is
associated with magnetic reconnection
\cite{1994MNRAS.270..480T,2005A&A...430....1G,2006A&A...457..763G,2012MNRAS.422.3092G}%
, neutron decay
\cite{1999ApJ...521..640D,2003ApJ...585L..19B,2006MNRAS.369.1797R,2007A&A...471..395K}%
, collisional heating \cite{2010MNRAS.407.1033B} and internal shocks
\cite{2002MNRAS.336.1271D,2010ApJ...725.1137L,2011MNRAS.415.1663T}.

Consider generic dissipation models following Ref.~\refcite{2005ApJ...628..847R}. When
the kinetic energy of the outflow is dissipated, electron-positron
pairs may be created and the pair photosphere where the optical depth of the
outflow due to Compton scattering on pairs reaches unity may emerge. When
continuous dissipation occurs in the steady coasting wind, the pair photosphere
can be found from the balance between the rate of energy dissipation and the
expansion rate $c\Gamma/r$. The former is given by $n_{\gamma}\sigma_{T}c$,
where $n_{\gamma}=L/(4\pi r^{2}m_{e}c^{3}\Gamma^{2})$ is the comoving photon
density, $m_{e}$\ is the electron mass, $L=\epsilon L_{0}$\ is the dissipated
radiative luminosity in the observer frame, $L_{0}$\ is the luminosity of the wind
and $\epsilon$ is the efficiency parameter. Assuming that photons have
sufficient energy to produce pairs one can obtain the photospheric radius of pairs produced by dissipation of energy as%
\begin{align}
R_{\pm,d}\lesssim\frac{L\sigma_{T}}{4\pi m_{e}c^{3}\Gamma^{2}}.
\end{align}
When compared to the second line in Eq.~(\ref{Rph}) we find $R_{\pm,d}\lesssim
R_{ph}m_{p}/m_{e}$. This pair photosphere may be located beyond the baryonic
one, thus changing the observed properties of the photospheric emission with
respect to the case without dissipation.

Another interesting effect is that the baryonic photosphere is also moved
further away after dissipation has occurred due to the decrease in the Lorentz factor.
This may be important in a scenario with sudden dissipation at a certain
radius above the saturation radius $\eta^{\frac{1}{a}}R_{0}$. When a
sufficiently large fraction of the kinetic energy is converted back into
thermal energy, the outflow may again become radiation-dominated and additional
acceleration will occur. In this case photospheric emission will necessarily
originate before the outflow starts to coast again. This means that impulsive
strong dissipation may change the character of the photosphere.

Two particular scenarios were analyzed in Ref.~\refcite{2006ApJ...642..995P}: with
slow heating \cite{1999ApJ...511L..93G} and with internal shock. While details
of these scenarios vary, in both cases the role of dissipation is two-fold.
Firstly, it converts a fraction of kinetic or electromagnetic energy of the
outflow into thermal energy. Secondly, it is supposed to supply additional
soft photons which may be Comptonized on the population of electrons
accelerated or heated by the dissipation process. Numerical simulations
performed in Ref.~\refcite{2006ApJ...642..995P}\ suggest that strong deviations from
pure black body spectra are possible for various parameters of these models.
In particular, when dissipation (e.g., shocks or heating) occur at high optical
depths of the outflow, the observed spectrum is dominated by the Wien
component which appears as result of Comptonization regardless of the values of the
parameters in the  models considered. For intermediate optical depths on the order
of unity, multiple scattering creates the flat energy per decade spectrum above the peak. For lower optical depths the dissipation is ineffective in
broadening the spectral component originating from the photosphere.

The basic feature of all photospheric models with dissipation is the
possibility to further broaden the spectrum of the photospheric emission. Besides, the intensity of the photospheric emission gets amplified. These models manage to reproduce some observed features in\ GRB spectra, see
e.g., Ref.~\refcite{2011MNRAS.415.1663T,2011MNRAS.415.3693R}. However, the need to
produce a sufficient number of photons for thermalization implies strong
constraints on the parameters of models involving dissipation, see e.g.,
Ref.~\refcite{2012MNRAS.422.3092G,2013ApJ...764..143V}. Moreover, so far there is no convincing dissipative model which on the one hand is based on a self-consistent relativistic hydrodynamics and/or kinetics, and on the other hand is able to reproduce many of the observational features of GRBs.

\section{Discussion and conclusion}

Photospheric emission is a natural ingredient of most popular models of GRBs
such as the fireball, fireshell or electromagnetic models. Despite the fact that GRBs are not generally characterized by a Planck
spectrum, recent observations demonstrate
that subdominant thermal components are likely to be present in many bursts.
This fact makes photospheric models of GRBs an attractive alternative to other
models involving only optically thin emission mechanisms.

In this review I have discussed a number of different physical aspects, in particular
relativistic effects, which make the photospheric emission from relativistic
outflows an interesting subject to study. From this analysis it follows that the expectation that a relativistic
photosphere produces a black body spectrum with a  Lorentz
boosted temperature is too naive. One can ask a general question: is there a way to
distinguish between a static optically thick source emitting from its
photosphere and a relativistically expanding one? It is the low energy part of the spectrum that holds the key to answering this question. The analysis performed so far implies that if the observational data are of
sufficiently good quality in order to resolve the low energy part of the
spectrum, the answer to this question is positive. As Fig.~\ref{photonthick}
clearly illustrates, the relativistically expanding source produces the
spectrum with a low energy slope different from the Planck case. Possible
dissipation near the photosphere may further broaden the observed spectrum and
make it similar to the one observed in many GRBs.

The photosphere is likely not the only source of radiation responsible 
for producing GRB emission, see Ref.~\refcite{2012ApJ...758L..34Z}. Due to relativistic
effects emission from the photosphere may arrive at the observer almost
simultaneously with the optically thin emission
\cite{2011MNRAS.415.3693R,2012MNRAS.420..468P,2013ApJ...763..125M}. What is
crucial, however, is that photospheric emission carries  unique information about
basic physical parameters of the relativistic outflow producing the GRB.

\vspace{0.5in}

{\bf Acknowledgement.} I am indebted to Alexey Aksenov, Vladimir Belinski, Carlo Luciano Bianco, Remo Ruffini and She-Sheng Xue for discussions. I would also like to thank Damien Begue, Robert Jantzen, and Ivan Siutsou for a critical reading of the manuscript.

\bibliographystyle{ws-ijmpd}


\end{document}